\documentclass[iop,revtex4]{emulateapj}
\usepackage{amsmath,amssymb}
\usepackage{graphicx}

\slugcomment{}

\shorttitle{Network Topology}
\shortauthors{Hong et al.}

\begin{document}


\title{Discriminating Topology in Galaxy Distributions using Network Analysis}


\author{ 
Sungryong Hong\altaffilmark{1,2}, 
Bruno Coutinho\altaffilmark{3},
Arjun Dey\altaffilmark{1}, 
Albert -L. Barab\'asi\altaffilmark{3,4,5}, 
Mark Vogelsberger\altaffilmark{6},
Lars Hernquist\altaffilmark{7},
and 
Karl Gebhardt\altaffilmark{2}
}

\altaffiltext{1}{National Optical Astronomy Observatory,  Tucson, AZ 85719, USA}
\altaffiltext{2}{Department of Astronomy, The University of Texas at Austin, 2515 Speedway, Stop C1400, Austin, TX 78712, USA}
\altaffiltext{3}{Center for Complex Network Research and Department of Physics, 
Northeastern University, Boston, Massachusetts 02115, USA}
\altaffiltext{4}{Department of Medicine and Channing Division of Network Medicine, Brigham and Women's Hospital, 
Harvard Medical School, Boston, Massachusetts 02115, USA}
\altaffiltext{5}{Center for Network Science, Central European University, 1051, Budapest, Hungary}
\altaffiltext{6}{Department of Physics, Kavli Institute for Astrophysics and Space Research, 
Massachusetts Institute of Technology, Cambridge, Massachusetts 02139, USA}
\altaffiltext{7}{Harvard-Smithsonian Center for Astrophysics, 60 Garden Street, Cambridge, Massachusetts 02138, USA}
\begin{abstract}
The large-scale distribution of galaxies is generally analyzed using
the two-point correlation function. However, this statistic does not
capture the topology of the distribution, and it is necessary to
resort to higher order correlations to break degeneracies. We
demonstrate that an alternate approach using network analysis can
discriminate between topologically different distributions that have
similar two-point correlations. We investigate two galaxy point
distributions, one produced by a cosmological simulation and the other
by a L\'evy walk, that have different topologies but yield the same
power-law two-point correlation function.  For the cosmological
simulation, we adopt the redshift $z = 0.58$ slice from Illustris
(Vogelsberger et al. 2014A) and select
galaxies with 
stellar masses greater than $10^8$$M_\odot$.
The two point correlation function of these simulated galaxies
follows a single power-law, $\xi(r) \sim r^{-1.5}$.  Then, we generate
L\'evy walks matching the correlation function and abundance with the
simulated galaxies.  We find that, while the two simulated galaxy point distributions 
have the same abundance and two point correlation
function, their spatial distributions are very different; most
prominently, \emph{filamentary structures}, which are present in
the simulation are absent in L\'evy
fractals.  To quantify these missing topologies, we adopt network
analysis tools and measure diameter, giant component, and transitivity
from networks built by a conventional friends-of-friends recipe with
various linking lengths.  Unlike the abundance and two point
correlation function, these network quantities reveal a clear
separation between the two simulated distributions; 
therefore, the galaxy distribution simulated by Illustris is not a L\'evy
fractal quantitatively. 
We find that the described network
quantities offer an efficient tool for discriminating topologies and
for comparing observed and theoretical distributions.  
\end{abstract}



 \section{Introduction}
 
Throughout the history of the Universe, various geometrical and
topological features have formed, evolved, and vanished in the cosmic
energy and matter distribution.  It is undeniably critical to quantify
and measure such features, since many of them can provide definitive
probes for constraining important cosmological parameters.

During the past two decades, studies of anisotropic features
in the cosmic microwave background (CMB),
specifically acoustic peaks, have motivated
the so-called $\Lambda$ cold dark matter ($\Lambda$CDM)
cosmology as a standard paradigm (e.g., Hinshaw et al. 2013, Aghanim
et al. 2015) and made a new step forward in precision cosmology.
Various experiments (Levi et al. 2013, Delubac et al. 2015, Zhao et
al. 2015), currently beginning or underway, are mapping out the
expansion history of the Universe with unprecedented accuracy, by
measuring baryon acoustic oscillations (BAO).  These experiments will
also result in the most detailed maps of the large-scale galaxy
distribution over a wide range in redshifts, from $z\sim0$ to
$z\sim3$.

The successes of measuring CMB acoustic peaks and BAO features
demonstrate how important the two-point correlation functions (or
power spectra) are for quantifying cosmic structures.  Higher order
$n-$point correlation statistics are essential for analyzing 
cosmic structures. For example, the three and four point correlation 
functions (or, bi- and tri-spectra) can constrain the
non-Gaussianity of primordial quantum fluctuations (Barkats et
al. 2014, Ade et al. 2015); however, these measures 
are computationally challenging
(e.g., Kulkarni et al. 2007, Gil-Mar\'in et al. 2015).


Along with the successful $n-$point statistics, many topological
measurements have been introduced, such as genus numbers and Minkowski
functionals (Gott, Weinberg \& Melott 1987, Eriksen et al. 2004).  To
identify voids and filaments, various methods have been adapted from
other fields of science, including minimum-spanning trees,
watersheds, Morse theory, wavelets, and smoothed Hessian matrices 
(e.g., Barrow, Bhavsar \& Sonoda 1985, Sheth et al. 2003, Mart\'inez et al. 2005,
Arag\'on-Calvo et al. 2007, Colberg 2007, Sousbie et al. 2008, Bond, Strauss \& Cen
2010, Lidz et al. 2010, Cautun et al. 2013).  While these topological diagnostics have
provided important insights into the nature of structure in the
Universe, this wide but heterogeneous range of applied methodologies
reflects how difficult it is to find a consistent and comprehensive
framework for quantifying and measuring the topology of the Universe,
in contrast to the successful $n-$point statistics.

Many of these studies generate a continuous density
field by smoothing the galaxy point distribution and then measuring geometric
topologies of genus numbers and Minkowski functionals.  Our approach, which 
we term ``network cosmology'', is to characterize the topology of the 
discrete point distribution directly using 
graph theory and network algorithms.

As a pilot study to explore new ways to quantify cosmic topologies,
Hong \& Dey (2015; hereafter, HD15) applied the analysis tools 
developed for the study of complex networks (e.g. Albert \& Barab\'asi 2002, Newman 2010)
to the study of the large-scale galaxy distribution. 
The basic
idea is to generate a graph (i.e., a ``network'') composed of vertices (nodes) and
edges (links) from a galaxy distribution, and then
measure network quantities used in graph theory. 
In this paper, we demonstrate the utility of these techniques for differentiating between point distributions that have identical two-point correlations but different spatial distributions and topologies. 

Our paper is organized as follows. In \S2, as a more specific
introduction to this paper, we offer a general discussion about what
types of features can be measured from galaxy survey data, the strong
and weak points of $n-$point statistics, and how network
representations of galaxy distributions can improve our ability to
quantify topological features in the Universe.  In \S3, we describe
our samples to be investigated, the snapshot of Illustris data
(Vogelsberger et al. 2014A) and L\'evy walks with various parameters.
In \S4, we present the two-point correlation functions and network
measurements from the samples and discuss the results.  Then, we
summarize in \S5.

\section{Geometric Configurations vs. Topological Textures in Galaxy Surveys}\label{sec:topologyintro}

Sections 2.1 and 2.2 present the definitions of geometry and topology
used in this paper and our overall philosophy in applying methods
of network analyses to galaxy distributions. Readers can skip these
sections without losing the main thread of this paper.

\subsection{Geometry vs. Topology}\label{subsec:comp}

The terms {\it geometry} and {\it topoplogy} are often used interchangeably in astronomical contexts. Geometry can be defined as the study of shapes of known metric dimensions, whereas topology refers to the intrinsic shape properties that are invariant to deformation (i.e., {\it homotopic}). For example, triangles are 3-sided geometric shapes that are characterized by the measures of their angles and sides. However, removing all metric features from triangles, we can also represent them topologically as a metric-free structure with three vertices where each vertext is connected to the other two by two edges. Another well-known example is the comparison between a mug and a donut; these are different geometric shapes with a common topology, the latter measured by a zero genus number.

Euler characteristics in graphs or genus numbers in manifolds are mathematically 
well-defined topological measures, invariant under homotopy or homeomorphism. 
However, most practical measurements are both geometric and topological, 
and do not have to be homotopy invariant in the strict mathematical sense
in order to be topologically meaningful. 
For example, the set of Roman alphabets is topological. We can consistently recognize letters irrespective of font or handwriting 
since each alphabet has its own distinct topology. 
``i'', ``k'', ``l''  
are topologically very different even in mathematically rigorous measures. 
However, ``i'' and ``j'' are indistinct topologically. They are discerned instead by the differences in length and curve (angle). The process of reading, i.e., visually measuring the characteristics of each letter, is predominantly topological but includes geometric aspects.

In galaxy surveys, $n-$point statistics are typical measurements, as presented in \S 1. 
These are geometrically driven measurements; $n-$point correlation functions contain 
specific information about distances and angles between galaxies. From 
a practical standpoint, this renders $n-$point statistics computationally 
challenging, since computation times are dominated by the handling of geometric
information.



If we are only interested in topological features, much of the geometric
information is redundant. 
For example, if we need to count all triangles in a friends-of-friends network from a certain galaxy distribution, 
we can run a network algorithm to count all triangular subgraphs. 
We do not need to measure the three point statistic for the problem of only counting triangles. 
Likewise, if we are interested in the number of holes for an object, 
we do not need to know whether it looks like a mug or a donut. 

Therefore, in practical analyses, 
we need to determine whether we are interested in quantifying \emph{geometric configurations}  
or \emph{topological textures}, when extracting measurements from galaxy survey data. 
Theoretically, a complete set of $n-$point statistics can suffice to characterize all aspects of a point distribution. However, such geometric analyses can 
be very inefficient when our prime focus is to quantify topological textures 
of the Universe. 


\subsection{Continuous Density Function vs. Discrete Point Distribution}\label{subsec:compdata}

To quantify topological structures of the Universe, 
many conventional studies have used geometric topology, 
where a metric topology is well-defined in a continuous cosmic matter 
distribution, $\rho(x)$, or its density contrast, $\delta(x)$.  
In this approach, discrete observables such as galaxy or halo distributions, $n(x)$, are considered as biased samplings 
of the underlying continuous cosmic matter distribution. 
Therefore, we generally smooth this discrete point distribution to approximate the continuous mass field.

In a different and empirical approach, we do not 
smooth over the discrete observable, $n(x)$. 
Instead, we build a network structure (or, a graph) from this discrete observable, 
and measure network quantities; 
hence, \emph{algebraic topology} from \emph{discrete observables}, contrast to the previous approach 
of \emph{metric topology} from \emph{continuous observables}.  
Hereafter, we refer to the former as ``DA" (discrete and algebraic) approach, 
and the latter as ``CM'' (continuous and metric) approach.

The CM and DA approaches differ in methodology. 
The CM approach is based on differential geometry and topology; 
hence, parameters of geometric shape and topology are derived from differentials or integrals of the density field.   
For example, 
the Hessian matrix is derived from partial differentials of the density field.
From the eigenvalues of this matrix, the clusters, walls, and filaments of the density field 
are classified (Arag\'on-Calvo et al. 2007, Bond et al. 2010, Cautun et al. 2013). 
Minkowski functionals are defined using integrals of the density field 
to quantify geometric and topological features such as area, perimeter, and genus 
(Mecke et al. 1994, Park et al. 2005, Hikage et al. 2008, Ducout et al. 2013).

On the other hand, our DA approach (which we refer to as ``network cosmology'') mostly utilizes network algorithms 
developed and used in computer science, mathematics, physics, and sociology. 
With 
its roots in Euler's brilliant solution to the K\"onigsburg bridge problem (Euler 1741), 
network science has grown rapidly during the last two decades, driven by the growth 
of computing power, large databases, and internet infrastructures 
(Albert \& Barab\'asi 2002, Barab\'asi 2009, Newman 2010). 
Networks can be constructed for studying subjects as diverse as the relationships between costarring actors, protein interactions, paper citations, the food web, power grids, traffic patterns, the world wide web (WWW), etc. 
Many network tools have been developed to extract useful information from these various kinds of big data networks. 
We have attempted, therefore, to utilize these network tools for investigating galaxy survey data. 
For example, \emph{PageRank} was developed for prioritizing the importance of WWW documents, 
used in the search engine, \emph{Google}\footnote{http://www.google.com} (Page et al. 1999).  
We can measure these \emph{PageRank}s for galaxies, once we build a galaxy network from galaxy survey data.
As we have a friend recommendation from \emph{Facebook}\footnote{http://www.facebook.com}, 
such a recommendation algorithm also can be applied to our galaxy network. 
This is the basic philosophy of our network cosmology. 


Early attempts of applying network science tools to galaxy point
distributions made use of percolation methods and the minimum
spanning tree, or MST (see, e.g, Shandarin et al. 1983AB, Barrow et al. 1985, Colberg 2007).
Since these pioneering papers, the tools developed for analyzing
networks have proliferated and mathematically matured. Our earlier
work, HD15, investigated galaxy distributions using various measures
of network centrality (degree, betweenness, and closeness). In this
paper, we apply the network measures of {\it diameter}, {\it giant
component fraction}, and {\it transitivity} to simulated galaxy
point distributions.

\subsection{Degeneracy in Two-point Correlation Function}

It has long been reported that observed galaxy populations exhibit single
power-law clusterings within several tens of megaparsecs
in comoving scale (e.g., Davis \& Peebles 1983, Shandarin \&
Zeldovich 1989, Adelberger et al. 2005).
Within the cold dark matter paradigm of galaxy formation, galaxies are
biased tracers of the underlying matter distribution and the
clustering properties of different galaxy populations can be diverse,
depending on how galaxies populate their dark
matter halos. Analysis of the two-point correlation function of
different galaxy populations has resulted in the idea of the ``halo
occupation distribution''; i.e., the probability that a given halo
contains a certain number of galaxies, and has given rise to various
analytic and probabilistic formulations of this function (Berlind \&
Weinberg 2002, Zheng et al. 2005, Tinker 2007).  From these halo
occupation studies, there should be a transitioning scale from the
dominance of the one halo term to the two halo term; hence
there is no need for galaxies to show a single, seamless power-law
clustering trend.  The apparent single power-law behavior, especially for
low redshift galaxies, is thought to be due to massive galaxy clusters
whose contribution erases the transition feature (Berlind \&
Weinberg 2002).

For a single power-law correlation, a couple of methods have been
proposed to generate mock galaxies including L\'evy walks (Mandelbrot
1975) and multi-layered shells akin to Russian dolls or onion rings
(Soneira \& Peebles 1978; hereafter, SP78).  These models are
``statistical'', since, unlike galaxies in simulations or the real
Universe, their clustering properties do not originate from
``gravitational'' interaction, but instead from a statistical fractal
realization.  These models can be tuned to match both the abundance
and the two-point correlation function of the observed galaxy
distribution.

Now, we raise two questions: 

(1) What is the gap between gravitational and statistical
realizations?, and

(2) Can we quantify the gap to finally test how much statistical
models are reliable as mocks?  

These are based on doubts about the sufficiency of information from
abundance and two-point statistics for testing cosmologies and
features that are missing in two-point statistics.

Interestingly, it is trivial for the human eye to capture the gap
between statistical models and observed (or simulated) galaxy
distributions.  SP78 reproduced a reasonable first approximation of
the observed galaxy distribution using their fractal model to
match the observed single power-law clustering.  And, due to some
visual gap in spatial distributions between their models and observed
galaxies, they remarked on the ability of the human eye, inherently
optimized to detect topological patterns rather than mathematical
geometries.  SP78 note that patterns easily discriminated by the human
eye are difficult to quantify, when compared to mathematically
straightforward $n-$point statistics.  Overall, SP78 implied that
there are some features, easily captured by the human eye, that are
not easily quantified by $n-$point statistics.

In this paper, we show that statistical ensembles produced by L\'evy
walks do not resemble simulated galaxies upon visual inspection (\S3),
agreeing with the same qualitative conclusion of SP78.  However, to
make a new step forward, we propose that topologically motivated
diagnostics, especially the network measurements adopted in what
follows, can quantify such eye-capturing features.

To test our proposal, we employ a simple setup, as follows.  First, we
adopt simulated galaxies as a cosmological sample.  While there are
discrepancies between observed and simulated galaxies, cosmological
hydrodynamic simulation can provide accurate three-dimensional
positions with realistic galactic properties, appropriate as a simple
pilot study without any observational complication.  Second, we
generate a statistical ensemble, using L\'evy walks, to match the
two-point correlation function of the simulated galaxies.  The next
section, \S3, will cover these two steps and present the spatial
distributions of the simulated and statistical models to show any
visual gap between them.  Finally, we measure network quantities for
the simulated sample and statistical ensemble in \S 4.  These network
measurements will explain why we recognize a difference between
statistical and gravitational realizations by sight, while they have
practically the same clustering property.  This will lead us to
discuss the limitations of $n-$point statistics and the potential of
network measurements as complementary topological diagnostics.

\section{Galaxy Distributions with Single Power-law Clustering}

In this section, we describe the two sets of simulated galaxy
distributions, one resulting from a hydrodynamic cosmological simulation and the
other a fractal generated by a simple L\'evy walk.

\subsection{Simulated Galaxies : Illustris Data}

The Illustris cosmological simulation is a modern, publicly available
simulation that computes the formation and evolution of both dark
matter and baryonic structures (Vogelsberger et al. 2014AB, Genel et
al. 2014, Nelson et al. 2015).  Illustris was performed using simple
models for the complex physics of star formation and growth of
supermassive black holes and associated feedback processes (Springel
\& Hernquist 2003; Springel et al. 2005; Di Matteo et al. 2005;
Sijacki et al. 2007; Vogelsberger et al. 2013). Illustris was run with the
moving mesh code Arepo (Springel 2010), an approach that offers
advantages in flexibility and accuracy compared to other methods
commonly used in cosmology (e.g. Vogelsberger et al. 2012, Keres et
al. 2012, Sijacki et al. 2012, Nelson et al. 2013, Zhu et al. 2015).
It yields one of the best simulated representations of galaxy
morphologies created within the context of a simulation, and hence
provides a reasonable dataset to mimic the properties of galaxies
resulting from an observational survey.

Specifically, we chose a single snapshot (\# 100) from the Illustris simulation.
Its corresponding redshift is 0.58, by which time
non-linear structures are well-developed; hence we observe rich
topological structures.  The size of the simulation box is $75
~h^{-1}$Mpc in comoving coordinates.  The resolution of the dark matter mass
is $6.26\times10^6$ M$_\odot$ and the resolution of the baryonic mass
$1.26\times10^6$ M$_\odot$.  
We selected galaxies with stellar mass $\ge 10^8 M_\odot$; this yields a sample of 75,050 galaxies. Hereafter, we
refer to this sample as ``Snap100''.
 
The top-left panel of Figure~\ref{fig:one} shows the two-dimensional
spatial distribution of Snap100, projected along the z-axis.  We can
identify rich structures of clusters and filaments.  The red-open
diamonds in the top-right panel of Figure~\ref{fig:one} show the two
point correlation function of Snap100, measured using the method from
Landy \& Szalay (1993).  We do not apply integral constraints to any
of the samples in this paper, since they are minor and contribute the
same amount due to the equal survey volume.  Power-law slopes can be
slightly shallower, when integral constraints are applied.  The
clustering of galaxies in Snap100 is well represented by a single
power-law with the slope, $\gamma \sim 1.5$.
 
\subsection{Statistical Fractal Galaxies : L\'evy Walks}

A L\'evy walk (or a L\'evy flight) is a random walk, whose step-size 
$l$ follows the distribution 
\begin{equation}\label{eq:lw}
P( > l)  = \left \{ \begin{array}{ll} 
(l_0/l)^\alpha & \textrm{for}~~ l \ge l_0 \\
1 & \textrm{for}~~ l < l_0, 
\end{array} \right.
\end{equation}
where $l_0$ is a minimum step-size and $\alpha$ is a fractal dimension. 
The L\'evy walk was introduced by Mandelbrot (1975) in cosmology as a method 
for generating a fractal galaxy distribution. 
The two-point correlation function of a L\'evy walk follows a power law 
\begin{eqnarray}
\xi(r) & = & C(l_0, \alpha) ~r^{-\gamma},   
\end{eqnarray}
where $\gamma = 3 - \alpha$ and $C(l_0, \alpha)$ is a constant determined by $l_0$ and $\alpha$ .

\subsubsection{Periodic Boundary Condition: L\'evy Walk in a Box}

The typical L\'evy walk presented above is an unbound random walk.  To
compare a L\'evy walk with the cosmological simulation, we need to confine
the walks to a cubic box.  Hence, we apply a periodic boundary
condition; we refer to this new L\'evy walk as ``L\'evy Walk in a
Box'' (LWIB).  The periodic boundary condition does not change the
slope of two point correlation function, $\gamma = 3 - \alpha$, but
affects the clustering amplitude, $C$,
\begin{eqnarray}
\xi(r) & = & C(l_0, \alpha, N_g, L) ~r^{-\gamma},   
\end{eqnarray}
where $N_g$ is the total number of L\'evy walks and $L$ is the size of
the box.  The newly entered walks resulting from
the periodic boundary condition
contribute as random encounters to the previously occupied
walks. Hence, when increasing $N_g$, the clustering amplitude, $C(l_0,
\alpha, N_g, L)$, decreases in LWIB models.  We set $N_g = 75,050$ and
$L = 75 ~h^{-1}$Mpc to compare LWIB with Snap100.

\subsubsection{Proximity Adjustment: Tweaking Small-Scale Clustering}

\begin{figure*}[t]
\centering
\includegraphics[height=8.5 in]{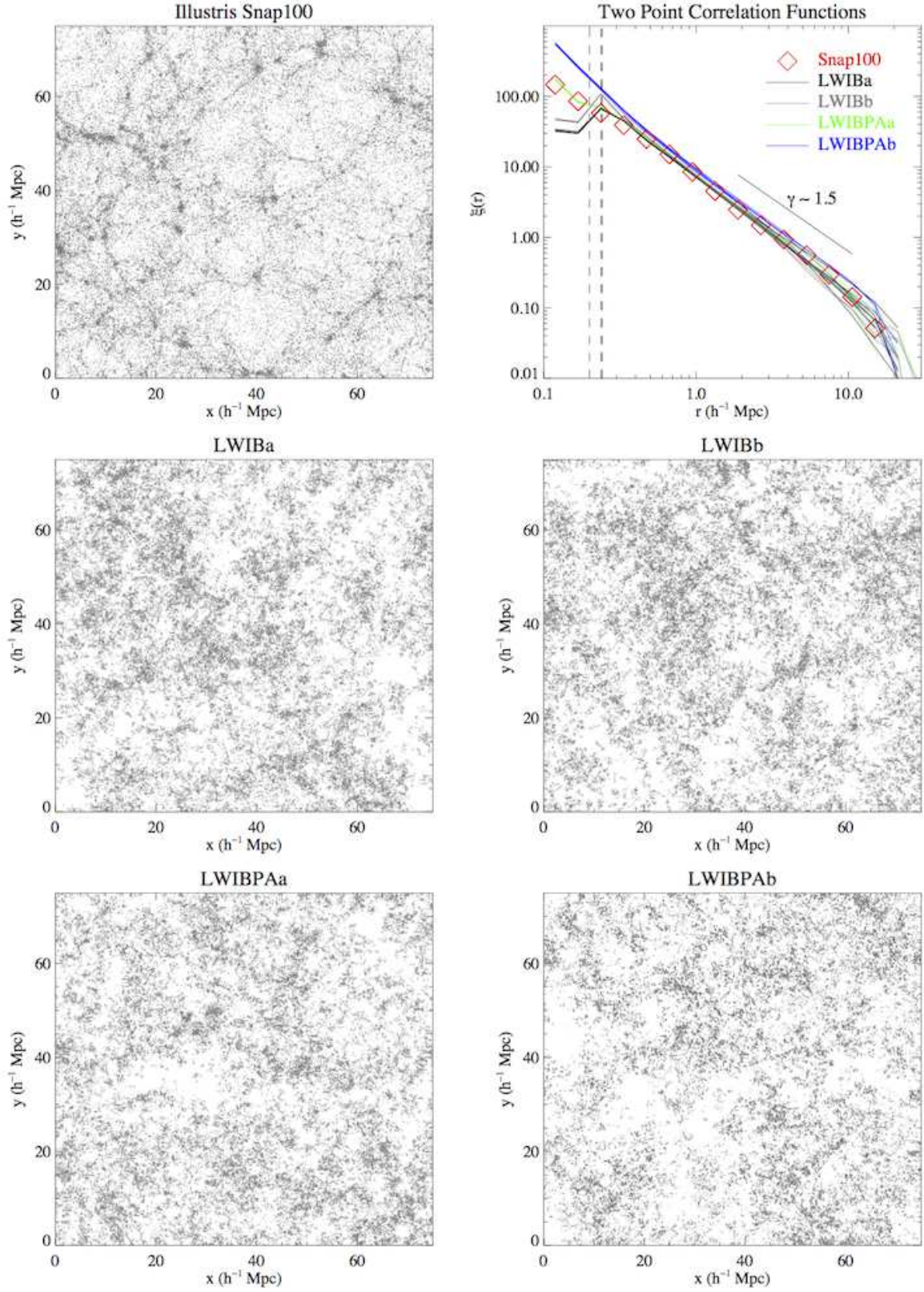}
\caption{The top-left panel shows the spatial distribution of  the Illustris galaxy distribution at $z = 0.58$ (``Snap100'') 
in comoving coordinates and the middle and bottom panels show L\'evy walk galaxies 
with various parameters, summarized in Table 1. 
The top-right panel shows the two point correlation function for each sample. 
The dashed vertical lines represent the minimum step sizes of L\'evy Walk models.  
We produce 100 realizations for each L\'evy walk and, here, we present 5 measurements (hence, 5 lines for each) 
to illustrate ensemble variances.     
The major difference, even clear in visual inspection, between the cosmological simulation and L\'evy fractals 
is the \emph{filamentary structure}, which is absent in the L\'evy fractal realizations. 
}\label{fig:one}
\end{figure*}

Although L\'evy walks are an elegant way to produce galaxies following
a single power-law clustering, the caveat is that their power-law
clustering property is only valid for $r>l_0$.  All galaxy pairs
closer than this minimum length are \emph{random encounters}
resulting in flat clustering for $r \leq l_0$.  The middle panels of
Figure~\ref{fig:one} show the spatial distributions of two LWIB
models, LWIBa and LWIBb; their model parameters are summarized in
Table 1.  The top-right panel shows the two-point correlation
functions of these two models, LWIBa (black) and LWIBb (grey).  The
vertical dashed lines represent the minimum step sizes: $l_0 = 0.2$
(grey) and $l_0 = 0.24$ (black).  The two LWIB models match well the
two point correlation function of Snap100 for $r > l_0$.  However, for
$r \leq l_0$, the clustering flatten out due to their intrinsic
limitations, as noted above.  To extend the power-law behavior to the
smaller scales $r \leq l_0$, we need to make those random close
pairs geometrically more compact.  We refer to this small-scale tuning of
clustering as ``Proximity Adjustment'' (PA).

There are many empirical approaches for determining the proximity
correction.  Our method is to require: (1) the correction to be based
on the LWIB, and identical to the latter in the limit of zero
correction; and (2) that the corrections only be applied on small
scales $r \le l_0$.  We refer to the models satisfying these two
criteria as ``L\'evy Walks in a Box with Proximity Adjustment"
(LWIBPA).

Specifically, we choose a simple extension of LWIB for our LWIBPA
model, as follows.  First, from the initial position (or the current
position), we generate the next walk by LWIB with $(l_0, \alpha)$.
Second, we find the nearest neighbor from the new walk position.
Third, if the distance from the nearest neighbor $r_{min}$ is larger
than the minimum step size $l_0$, i.e., $r_{min} > l_0$, then we
accept this walk and proceed to the next iteration.  If $r_{min} \leq
l_0$, we calculate a new step size from the new power-law of $(l_m,
\beta)$, where $l_m < l_0$.  If this new step size, $r_{new}$, is
larger than $l_0$, i.e., $r_{new} > l_0$, then we discard this PA
process to accept the original LWIB position and proceed to the next
iteration. If $r_{new} \leq l_0$, we take a random roll, $\mu \in [0,
1)$. If this roll, $\mu$, is larger than our acceptance threshold,
$p_\theta$, i.e., $\mu > p_\theta$, we again discard this PA process
to accept the original LWIB position and proceed to the next
iteration. Finally, for $\mu \leq p_\theta$ along with the previous
$r_{min} \leq l_0$ and $r_{new} \leq l_0$, we accept this PA
correction. We keep the direction between the nearest neighbor and
new walk position to only replace $r_{min}$ with $r_{new}$.  Whether
the PA correction is accepted or not, the next walk is calculated
based on the original LWIB position.  We build this PA recipe in a
conservative manner to keep the new LWIBPA as close as possible
to the original LWIB.  To briefly summarize, our LWIBPA model is a
broken two-power-law model with a threshold of acceptance
probability determining the choice between the two power laws, and
is represented by the five parameters $(l_0, \alpha, l_m, \beta,
p_\theta)$.

The bottom-panels of Figure~\ref{fig:one} show the spatial
distributions of our two LWIBPA models, LWIBPAa and LWIBPAb, where
their parameters are summarized in Table 1.  The top-right panel
shows the two point correlation functions of LWIBPAa (green) and
LWIBPAb (blue).  These two models are variants of the original LWIBa
model, sharing the same parameters $(l_0 = 0.24, \alpha = 1.5)$; but
having different acceptance probabilities, $p_\theta = 0.35$ for
LWIBPAa and $p_\theta = 1$ for LWIBPAb.  LWIBa also corresponds to the
model of $p_\theta = 0$.

\begin{deluxetable}{lccccc}
\tabletypesize{\scriptsize}
\tablecaption{L\'evy Walk Models\label{tbl-1}}
\tablewidth{0pt}
\tablehead{
\colhead{Name} & \colhead{$l_0$} & \colhead{$\alpha$} & \colhead{$l_m$} & \colhead{$\beta$} & \colhead{$p_{\theta}$}  
}
\startdata
LWIBa & 0.24 & 1.5 & -- & -- & --\\
LWIBb & 0.20 & 1.6 & -- & -- & --\\
LWIBPAa & 0.24 & 1.5 & 0.01 & 1.5 & 0.35 \\
LWIBPAb & 0.24 & 1.5 & 0.01 & 1.5 & 1.00\\
\enddata
\tablecomments{$l_0$ and $\alpha$ are the basic L\'evy walk parameters presented in Equation~\ref{eq:lw}. 
The others are for the ``proximity adjustment'' as explained in the text. 
None of these L\'evy walk models properly mimic the spatial distribution of Snap100. }
\end{deluxetable}

\section{Results}

\subsection{Missing Topologies in Two Point Correlation Function}

In the previous section, we described our adopted simulation sets,
Snap100, and L\'evy walk recipes, and measured their two-point
correlation functions, as summarized in Figure~\ref{fig:one}.

LWIBPAa yields a good match to the correlation function of Snap100, at
least to the accuracy of practical clustering studies.  LWIBa and
LWIBPAb can be considered, respectively, as ``lower'' and ``upper''
bounds of correlation functions encompassing suppressed and enhanced
small scale clustering.  LWIBb is a model with slightly different
parameters, $l_0 = 0.2$ and $\alpha = 1.6$, from LWIBa, demonstrating
that the clustering properties of LWIB models do not change abruptly
by choosing parameters nearby.  Hence, the four types of L\'evy walk
models illustrated in Figure~\ref{fig:one} span a good range of
possible L\'evy fractals, comparable to Snap100.

The important point is that while the L\'evy walk distributions
reproduce the two-point correlation function of the galaxy
distribution in the Illustris simulation, none of them mimic the
actual spatial distribution of the Snap100 galaxies.  In particular,
the L\'evy walks fail to reproduce the filamentary structure that is
so characteristic of actual galaxy distributions.  This implies that
L\'evy fractals are not appropriate for explaining the structure of
the (simulated) Universe, and two-point statistics are highly
degenerate.  As noted in \S \ref{sec:topologyintro}, this is because
topological features are elusive in $n$-point statistics, while human
eyes are more adapted to effectively recognize topologies of patterns
and connectivities.  Unlike what has been believed up to now, that
such eye capturing features are hard to quantify, in the following
sections we describe how such topologies can be measured using network
science tools.

\begin{figure}[t]
\centering
\includegraphics[height=7.0 in]{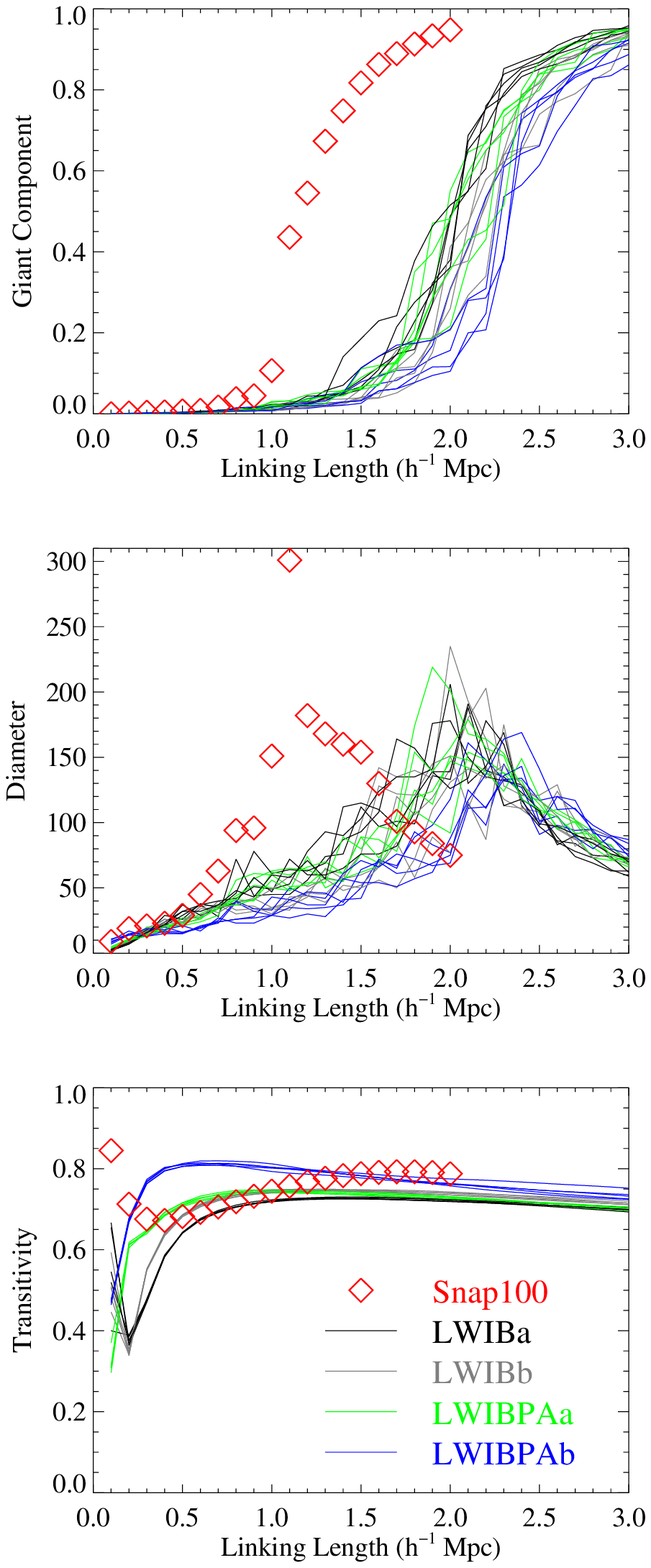}
\caption{ The three network measurements, giant component fraction (top), diameter (middle), 
and transitivity (bottom) vs. linking length for the five models, the Illustris $z = 0.58$ snapshot 
(``Snap100'') (red-open diamonds), 
LWIBa (black lines), LWIBb (grey lines), LWIBPAa (green lines), and LWIBPAb (blue lines).
For each L\'evy walk model, we plot 5 lines for 5 realizations like 
in Figure~\ref{fig:one} to illustrate statistical variances. 
All the three network measurements show clear separations between Snap100 and L\'evy fractals,
implying that L\'evy walk models fail to match the topological properties of Snap100. 
The galaxy distribution in the Illustris simulation is clearly not a L\'evy fractal.
}\label{fig:two}
\end{figure}

\begin{figure*}[t]
\centering
\includegraphics[height=6.0 in]{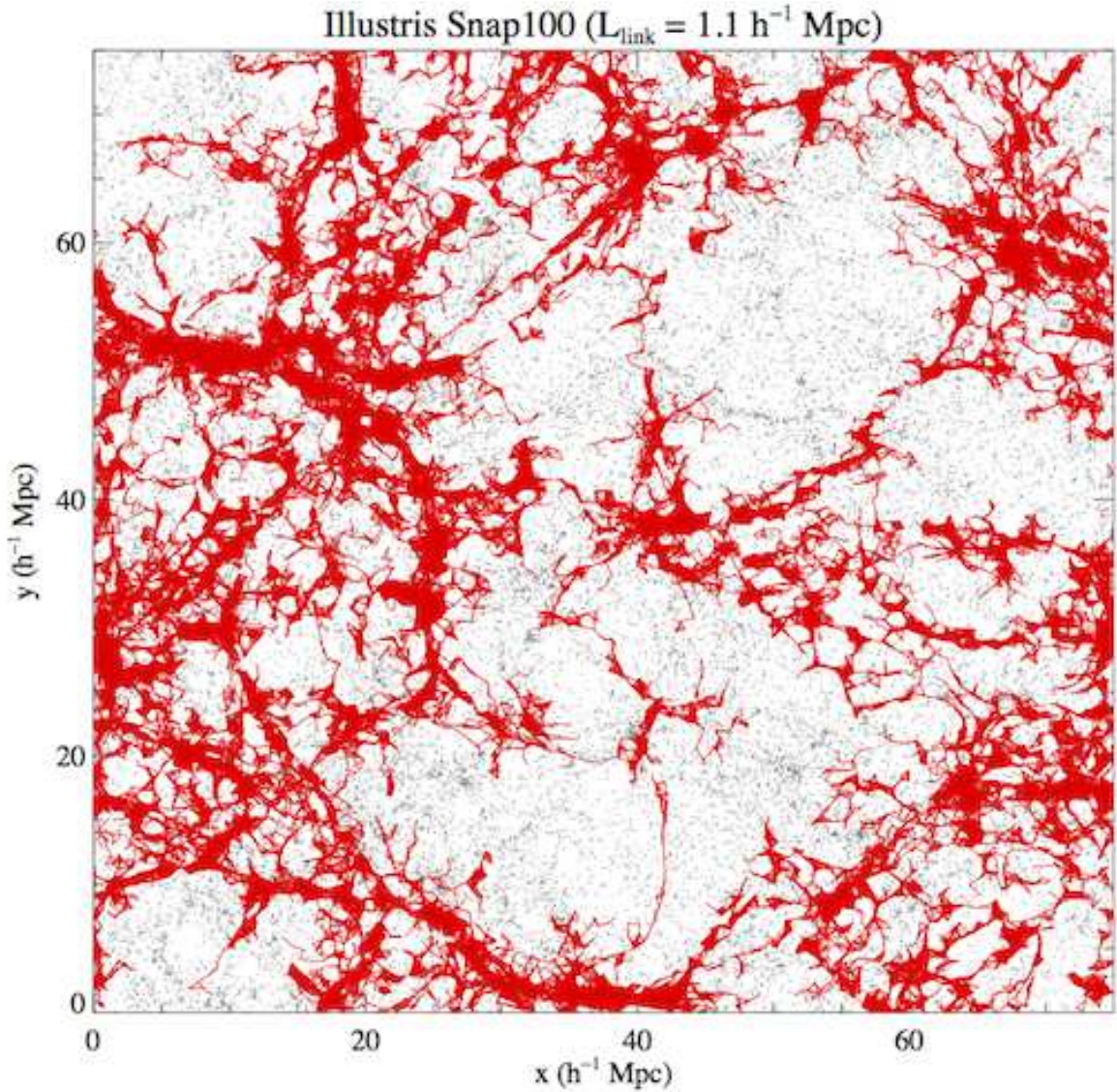}
\caption{ The simulated galaxies of the Illustris $z = 0.58$ snapshot 
(``Snap100''; grey dots) and edges connecting galaxies in the giant component (red line), 
visualizing the spatial network structure of 
the giant component.  The linking length is $ 1.1~h^{-1}$Mpc, where the diameter is maximized. 
The texture of this Snap100 giant component can be described as ``thin, diversifying, and filamentary''. 
}\label{fig:three}
\end{figure*}  

\begin{figure*}[t]
\centering
\includegraphics[height=6.0 in]{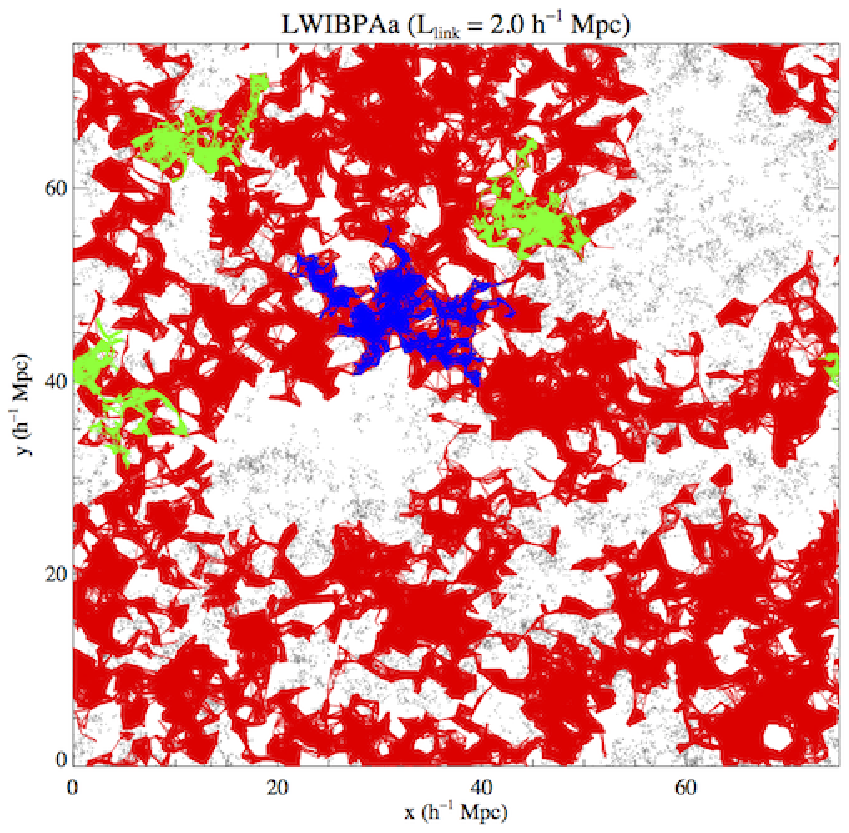}
\caption{ The same as Figure ~\ref{fig:three} 
but for LWIBPAa. The linking length is $ 2.0~h^{-1}$Mpc, where the diameter for LWIBPAa is maximized.
The texture of LWIBPAa's giant component (red lines) can be described as ``thick, clumpy, and modularized''. 
The blue lines show the edges of the
giant component (hence, the largest component), and the green lines of the second, third, and fourth largest components, 
for the linking length $ 1.1~h^{-1}$Mpc, comparable to Figure ~\ref{fig:three}. 
While the giant component of Snap100 shows a fully developed global structure at $ 1.1~h^{-1}$Mpc, 
the giant component (blue lines) and next largest components (green lines) 
of LWIBPAa are still localized due to the lack of topological bridges. 
}\label{fig:four}
\end{figure*}

\subsection{Network Analysis: Quantifying Missing Topologies}

A \emph{network} is a data structure composed of ``vertices'' (or
nodes) connected by ``edges'' (or links); also known as a \emph{graph}
in mathematics.  In the 21st century, network science (or graph
theory) has becomes one of the most critical tools in various fields, such as
bioinformatics, computer science, physics, and sociology.  In a
previous work (HD15), we explored the use of network measures
(betweenness, closeness and degree) to investigate the relationships
between galaxy properties and topology.  The results were promising,
but limited by the use of photometric (rather than spectroscopic)
redshifts to characterize the 3-dimensional galaxy distribution.
Readers interested in networks are referred to Newman (2003),
Dorogovtsev \& Goltsev (2008), Barth\'elemy (2011), and HD15 for
further information.

\subsubsection{Linking Length and Friends-of-friends Network}

To build a network from a given galaxy population,
we adopt the conventional friends-of-friends (FOF) recipe (e.g. Huchra \& Geller 1984, More et al. 2011, HD15).
 
For a given linking length $l$, we define the \emph{adjacency matrix} as, 
\begin{equation}\label{eq:lldef}
A_{ij} = \left\{ \begin{array}{ll}
	1 & \textrm{   if   } r_{ij} \le l,  \\
	0 & \textrm{   otherwise, } 
	\end{array} \right.
\end{equation}
where $r_{ij}$ is the distance between the two vertices, $i$ and $j$. 
This binary matrix quantitatively represents the
network connectivities of the FOF recipe. 
Many important network measures are derived from this matrix. 

\subsubsection{Network Topologies : Diameter, Giant Component, and Transitivity}

We measure three simple scalar
quantities, \emph{diameter, giant component, and transitivity,} 
from FOF networks for various linking lengths, using the open network library, 
\emph{igraph} (Csardi \& Nepusz 2006).
Due to their simple definitions, these measures are computationally cheap and widely used in complex networks. 
The question is whether these network science tools can
quantify the topological differences missed by two point statistics.

\emph{The Diameter} is the largest path length of shortest-pathways
from all pairs in a network.  The path length is defined as the number
of steps to reach from a certain vertex, $i$, to another, $j$.  Hence,
the pathways of minimum path length are the shortest pathways between
the vertices, $i$ and $j$; generally, there can be multiple shortest
pathways between a pair in an unweighted network.  When a linking
length is quite small to isolate all galaxies alone, the diameter is
trivially 0.  As the linking length is increased, the diameter grows
to reach a certain maximum value.  Since, for a very large linking
length, all pairs are connected by a single direct edge (in the
mathematical terms, forming a ``complete graph''), the diameter
asymptotically decreases to 1, after reaching the maximum value.
Hence, this varying curve of \emph{Diameter vs. Linking Length} is a
quantified topology, depending on pathway structures.

\emph{The Giant component} is the largest connected subgraph in a
network. As in the case of diameter, giant components are trivial for
the two extreme linking lengths.  For a small linking length that
isolates individual galaxies, the size of the giant component is 1.
In the opposite case of a very large linking length forming a complete
graph, the giant component size is equal to the total number of
vertices (galaxies).  Hence, the ratio of the size of the giant
component to the total number of vertices is a fraction that increases
from 0 to 1 monotonically with the linking length.  The rate of growth
of this ratio depends on topology; if a network has some topological
structures to connect vertices more efficiently, the fraction of the
giant component grows faster to reach 1 at a smaller linking length.

\emph{Transitivity} can be described as a ``triangle density'' for a
network.  It is defined as:
\begin{eqnarray}\label{eq:trans}
C & = & \frac{\textrm{number of closed paths of length two}}{\textrm{number of paths of length two}},   
\end{eqnarray}
where $C$ denotes the transitivity (Newman 2010).  A path of length
two means a ``$\vee$'' shaped connection; i.e. my friend-of-friend
configuration in a social network.  If my friend-of-friend is my
direct friend, this path of length two forms a closed path of length
two; i.e. a triangle ``$\triangledown$''.  Therefore,
Equation~\ref{eq:trans} predicts a higher transitivity value if there
are more
triangles in a network.  To some extent, transitivity can be
considered as a minimal (and topological) version of the three point
correlation function.

Figure~\ref{fig:two} shows the three network quantities for our 5
samples, Snap100 (red-open diamonds), LWIBa (black lines), LWIBb (grey
lines), LWIBPAa (green lines), and LWIBPAb (blue lines).  As in Figure
1, for each L\'evy walk model we plot 5 lines for 5 realizations to
illustrate statistical variance.  The three network quantities
are uniquely determined for a given linking length.  Namely, the three
plots of diameter, giant component, and transitivity vs. linking
length are self-consistently determined for a given spatial
distribution like $n-$point statistics without any further parameter
or assumption, except for their independent variable, linking length.


The top panel of Figure~\ref{fig:two} shows the results for giant
component fractions.  Now we can quantitatively discern Snap100
(red-open diamonds) from LWIBPAa (green lines), though they have
(practically) the same abundance and two-point correlation function,
shown in Figure~\ref{fig:one}.  All the other L\'evy walk models also
fail to match the growth curve of giant component fractions.  When
considering that LWIBa and LWIBPAb are, respectively, lower and upper
bounds of the small-scale clustering for LWIBPAa (and Snap100), the
failures of all L\'evy walk models to match giant component fractions
imply the fundamental difference in the pathway topology between
Snap100 and L\'evy fractals; Snap100 has more efficient pathways to
connect all galaxies at a shorter linking length than L\'evy fractals.
Very likely, this is due to the \emph{filamentary structures} in
Snap100, lacking in L\'evy fractals.

The middle panel of Figure~\ref{fig:two} shows the diameters.  We can
again see clear separations of Snap100 from the L\'evy walk models.
Snap100 reaches the maximum diameter, 300, at the linking length,
1.1$h^{-1}$Mpc, while L\'evy walk models reach the maximum diameters
around 200 for linking lengths near 2.0$h^{-1}$Mpc.  Even for 100
L\'evy walk realizations, none of the L\'evy walk models can match the
diameter measurements of Snap100.  Hence, both the size of the giant
component and diameter are network measures that discriminate the
L\'evy walk topologies from the Illustris simulation, despite the data
sets being constructed to have matching abundance and two-point
correlation statistics.

The linking length for maximum diameter  is  related to the inflection point of the
growth curve of giant component fractions; 
the rate of growth of the giant component decreases after reaching the maximum diameter. 
This transitioning feature occurs due to the ``saturation'' of connecting edges. 
At first (i.e., small linking length values), increasing the linking length results in adding new vertices and increasing the size of the connected network components. However, once the largest diameter is reached, increasing the linking length tends to form new pathways within the existing structure between more far-flung members and only slowly increases the overall size of the connected structure.
Therefore, the diameter is maximized at this critical scale, transitioning from ``growing phase'' to ``saturating phase''.
The previous percolation studies are closely related to this maximum diameter scale, 
though they have not measured these specific diameters.  
If the system size is infinite, the diameter measurements transit from finite values to an infinity near this scale.

The bottom panel of Figure~\ref{fig:two} shows transitivity results.
Again, none of L\'evy walk models mimic the transitivity curve of
Snap100.  We note that the statistical variances of transitivity
measurements are much smaller than the other measurements as shown in
Figure~\ref{fig:two}, since a single realization of the network is
statistically large enough for counting triangles.  Hence, the
difference of transitivities between Snap100 and L\'evy fractals also
suggests that Snap100 is topologically very different from L\'evy
fractals.

An interesting feature is the difference of convexities between
Snap100 (concave or ``cup'') and L\'evy fractals (convex or ``hat'').
The transitivities of Snap100 are high for small linking lengths, then
decrease to a minimum transitivity at 0.4 $h^{-1}$Mpc as the linking
length increases.  After this, the transitivities slowly increase to
0.8.  This transitivity trend of Snap100 is related to the transition
between the one-halo term to the two-halo term in halo occupation
clustering models (Berlind \& Weinberg 2002).  For small linking
lengths, most triangles form in cluster environments reflecting halo
substructures.  Hence, these ``intra-halo triangles'' (i.e., triangles
lying within one halo) dominate the transitivities for small linking
lengths, and result in a decreasing trend from a very high
transitivity.  On the other hand, for sufficiently large linking
lengths, ``inter-halo triangles'' (i.e., halo-halo-halo triangles)
dominate over intra-halo triangles, simply because their
configurations are more frequently found.  Since a ``$\vee$'' shaped
configuration becomes a triangle for a larger increased linking
length, the transitivities for inter-halo scales are generally an
increasing function.  Therefore, this is potentially a very
interesting point.

For L\'evy walk models, the origins of triangles are different from
Snap100.  For scales smaller than the minimum L\'evy walk step $l_0$
(0.2 $h^{-1}$Mpc and 0.24 $h^{-1}$Mpc in Table 1), the triangles
originate from ``random'' encounters or our ``proximity adjustment''
recipe.  For scales larger than the minimum L\'evy walk steps, the
fractal L\'evy walks shape the rest of the triangles.  Hence,
discontinuities occur at these breaking scales in transitivity curves.
The typical fractal transitivities increase to reach maximum values,
and then asymptotically decrease to around 0.75. These convex (or
``hat'') trends contrast to the concave (or ``cup'') shape of Snap100.

Figure~\ref{fig:three} shows the edges (red lines) connecting galaxies
in the giant component of Snap100, visualizing the spatial network
structure of the giant component.  The linking length is
1.1$h^{-1}$Mpc, where the diameter is maximized.  The texture of this
Snap100 giant component can be described as ``thin, diversifying, and
filamentary''.  Figure~\ref{fig:four} shows the same as
Figure~\ref{fig:three} for LWIBPAa.  The linking length is 2.0
$h^{-1}$Mpc, where the diameter for LWIBPAa is maximized.  The texture
of LWIBPAa's giant component (red lines) can be described as ``thick,
clumpy, and modularized''.  The blue lines show the edges of giant
component for the linking length 1.1 $h^{-1}$Mpc, comparable to Figure
~\ref{fig:three}.  While the giant component of Snap100 shows a fully
developed global structure at 1.1 $h^{-1}$Mpc, the giant component of
LWIBPAa (blue lines) is still localized due to the lack of topological
bridges.  Overall, the structural and topological differences between
Snap100 and L\'evy fractals are well reflected in network structure.

\begin{figure}[t]
\centering
\includegraphics[height=3.0 in]{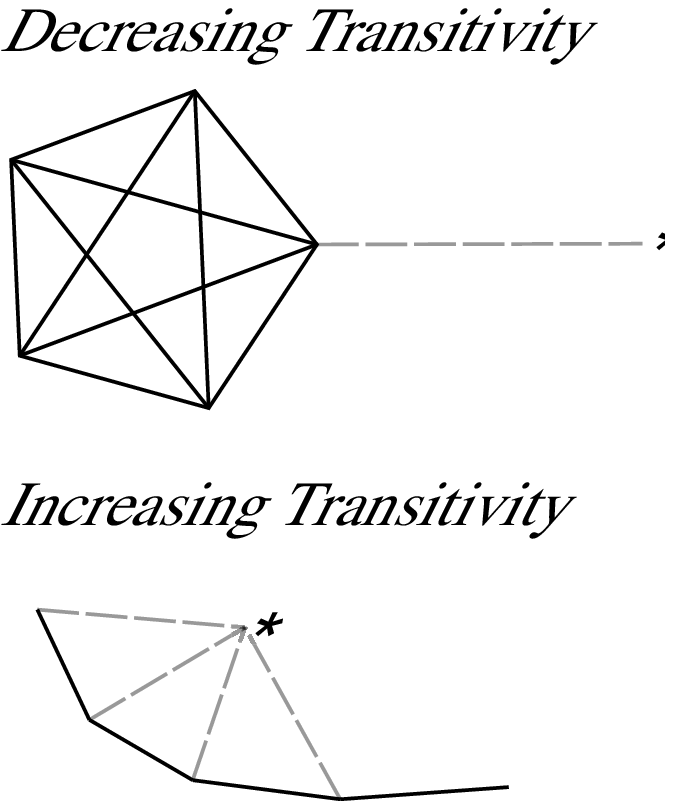}
\caption{Schemas demonstrating two possible cases to increase (top) and to decrease (bottom) transitivity values  
by adding a new vertex (asterisk) and its edges (dashed-grey lines).  In the top diagram, the new dashed edge 
produces 4 additional ``$\vee$'' configurations, but none of them form triangles. 
On the other hand, in the bottom diagram, three triangles form by the new vertex and edges. 
Hence, the different convexities of transitivity curves for Snap100 and L\'evy fractals 
reflect the intrinsic difference of topological structures.  
}\label{fig:five}
\end{figure}

Figure~\ref{fig:five} presents two basic schemas to demonstrate which
topological configuration can increase (or decrease) the
transitivity. We note that variation of transitivity depends on very
complex topological structures.  The schemas are only two possible
cases among many.

The top diagram of Figure~\ref{fig:five} shows that the new vertex
(asterisk) and edges (grey dashed lines) produce four additional
``$\vee$'' configurations, but none of them form a triangle; hence,
transitivity decreases by this new vertex.  This schema provides a
possible illustration as to why Snap100 shows a decreasing
transitivity trend at intra-halo scales.  On the other hand, the
bottom diagram of Figure~\ref{fig:five} shows that the new vertex and
edges form three additional triangles to increase the
transitivity. Basically a linear chain of walks is less efficient in
forming triangles than a gravitational pull to pack galaxies. This
explains why L\'evy walks show smaller transitivities at small scales
than Snap100.  However, such low transitivity values can be restored
as the linking length increases as in the bottom diagram.  Hence, the
different behaviors of transitivity between Snap100 and L\'evy
fractals reflect the different topological bindings of galaxies (or
walks); i.e., gravitationally packed solid ball vs.  linearly tangled
ball.

\section{Summary}

In this paper we have used a network approach to compare two galaxy
distributions with similar two-point correlation statistics but
different topologies, one derived from a cosmological simulation and
the other from a L\'evy walk.  
The network measures are computed directly from the point distribution 
of the galaxies, unlike past measures that characterize a smoothed continuous 
version 
of the point distribution. 
We find that the simulated galaxies and
L\'evy walks are statistically different in diameter, giant component,
and transitivity measurements, which shows that L\'evy walks fail to
mimic the topologies of the distribution of the simulated galaxies,
though they successfully match the abundance and two point correlation
function.

This implies that quantified topologies are important for testing
cosmologies.  While $n$-point statistics are undeniably useful
diagnostics, their topological complementaries are necessary to
properly test cosmologies and to prevent misinterpretation that
could result from over-simplified false-positive models.
 

\acknowledgments
We are grateful to an anonymous referee for comments that have improved this paper. 
SH's research activities have been supported by the National Optical Astronomy Observatory (NOAO) 
and the University of Texas at Austin, and AD's by NOAO.  
NOAO is operated by the Association of Universities for Research in Astronomy (AURA) under 
cooperative agreement with the National Science Foundation.  LH is
supported by NASA ATP Award NNX12AC67G and NSF grant AST-1312095.

\end{document}